\documentclass{DISproc}

\usepackage{caption,subcaption}

\newcommand{\herwig}{\textsc{Herwig}}
\newcommand{\pythia}{\textsc{Pythia}}
\newcommand{\sherpa}{\textsc{Sherpa}}
\newcommand{\lambdainit}{\lambda_{\rm init}}
\newcommand{\lambdafinal}{\lambda_{\rm final}}
\newcommand{\deltaif}{\Delta_{\rm if}}

\makeatletter
\def\localnumbers#1{\gdef\@localnumbers{#1}}                                  
\def\ps@localnumbers{\let\@mkboth\@gobbletwo
     \def\@oddhead{\hfil \@localnumbers}%
     \def\@oddfoot{{\slshape \@acronym}\hfil\thepage}%
     }
\makeatother

\begin{document}

\localnumbers{\parbox[t]{3.5cm}{KA-TP-24-2012\\ MAN/HEP/2012/07}}

\title{The Underlying Event in Herwig++}

\author{{\slshape Stefan Gieseke$^1$, Christian R\"{o}hr$^1$\footnote{Speaker},
Andrzej Si\'odmok$^{1,2}$}\\[1ex]
$^1$Karlsruhe Institute of Technology (KIT), 76128 Karlsruhe, Germany\\
$^2$The University of Manchester, Manchester, United Kingdom}

\contribID{xy}

\doi  

\maketitle

\thispagestyle{localnumbers}

\begin{abstract}
  We review the modelling of multiple interactions in the event generator
  \herwig{}++ and study implications of recent tuning efforts to LHC data. A
  crucial ingredient to a successful description of minimum-bias and
  underlying-event observables is a model for colour reconnection. Improvements
  to this model, inspired by statistical physics, are presented.
\end{abstract}

\section{Introduction}

Multiple partonic interactions (MPI) are vital for a successful description of
the underlying event (UE) in hard hadronic collisions and of minimum-bias (MB)
data from the Tevatron and the Large Hadron Collider (LHC).  A model of
independent multiple partonic interactions was first implemented in \pythia{}
\cite{Sjostrand:1987su}, where its relevance for a description of hadron
collider data was immediately shown. Meanwhile, all major event generators for
LHC physics, \herwig{}~\cite{Bmboxahr2008}, \pythia{}~\cite{Sjostrand:2006za,
Sjostrand:2007gs} and \sherpa{}~\cite{Gleisberg:2008ta}, contain MPI models.
The core MPI model in \herwig{}++, which is similar to the \textsc{Jimmy}
add-on~\cite{Butterworth1996} to the Fortran version of \herwig{}, was
introduced in Ref.~\cite{Bahr:2008dy}.  Additional hard parton-parton scatters
unitarize the hard jet cross section.  Also the jet-like structure of the
underlying event is reproduced by this model.  With soft components in multiple
parton interactions included, which is described in Ref.~\cite{Bahr:2009ek},
this model is sufficient to describe the UE data collected at the Tevatron.
First MB data from ATLAS \cite{Aad:2010rd}, however, e.g.~the pseudorapidity
distribution of charged particles, cannot be reproduced with the core MPI model
discussed so far.

As shown in Ref.~\cite{Gieseke:CRmodel}, which we summarize in this work, we can
significantly improve the description of MB and UE data from the LHC if we
include a model for colour reconnections (CR). The idea of CR is based on colour
preconfinement \cite{Amati:1979fg}, which implies that parton jets emerging from
different partonic interactions are colour-connected if they overlap in momentum
space. As the core MPI model does not take that into account, those colour
connections have to be adapted afterwards by means of a CR procedure.

\begin{figure}[t]
  \begin{minipage}[t]{.42\linewidth}
    \centering
    \includegraphics[width=\textwidth]{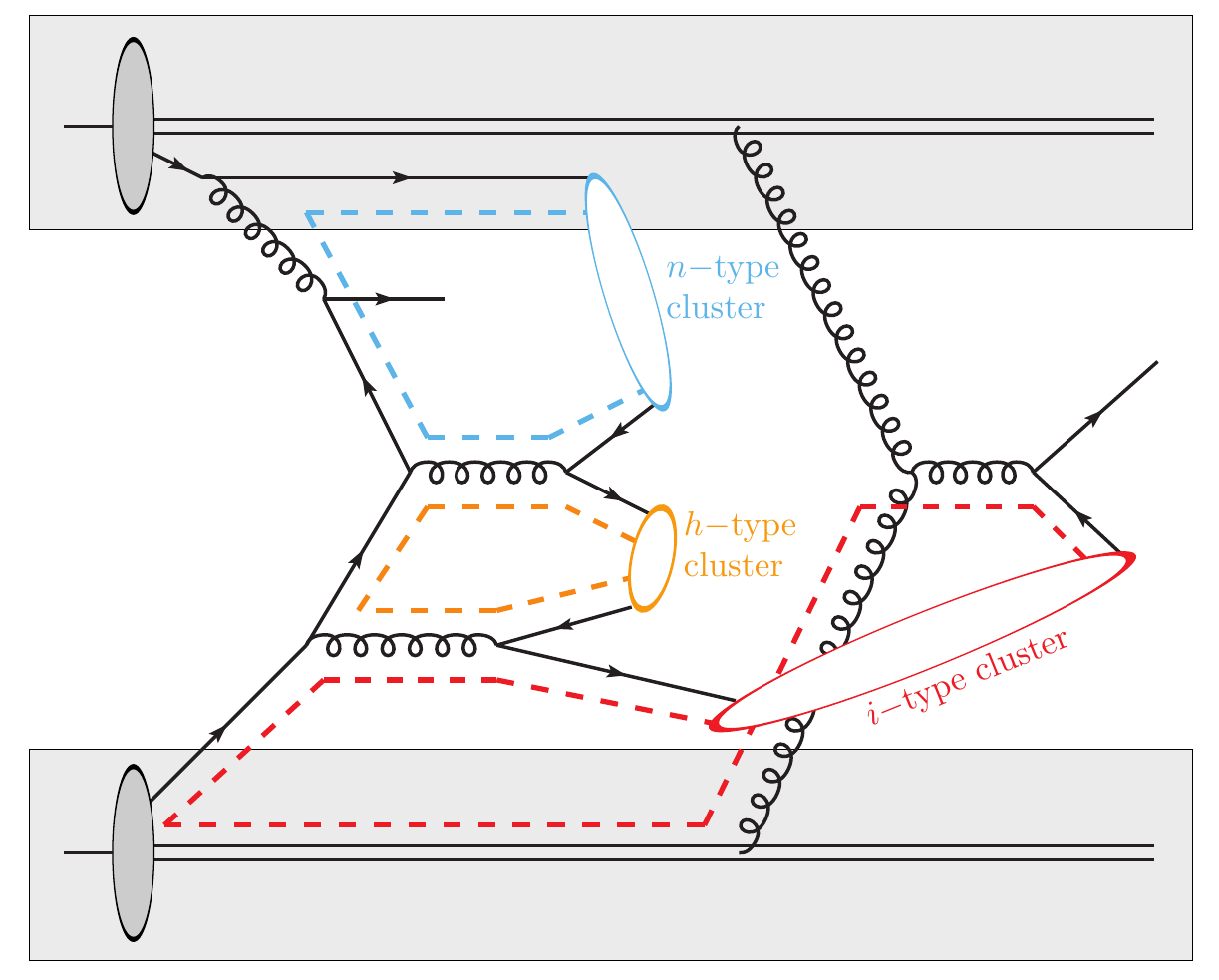}%
    \subcaption{Sketch of cluster classes in a simple $pp$ event. The classes
    are defined in the text.}
    \label{fig:classesdef}
  \end{minipage}%
  \hfill
  \begin{minipage}[t]{.54\linewidth}
    \centering
    \includegraphics[width=\textwidth]{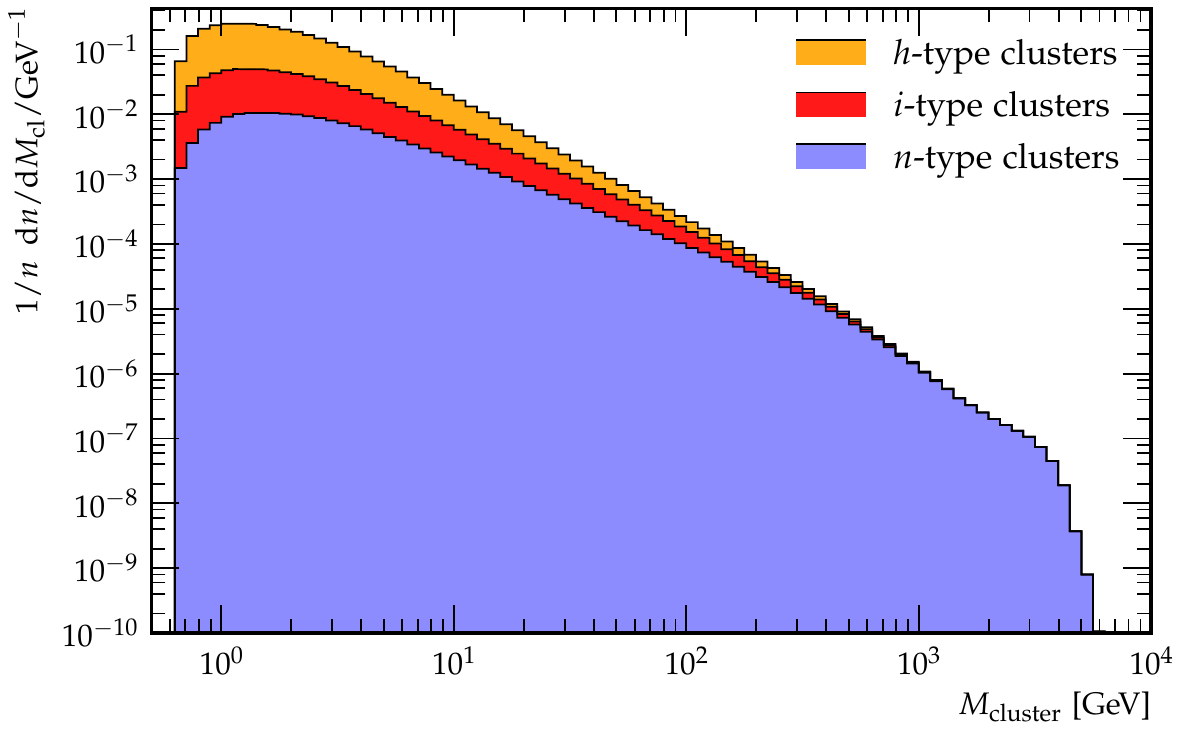}%
    \subcaption{Contributions of the cluster classes to the invariant cluster
    mass distribution in soft-inclusive $pp$ collisions at \unit{7}{\TeV}.}
    \label{fig:classesmass}
  \end{minipage}
  \caption{Classification of clusters in hadron collision events.}
  \label{fig:classes}
\end{figure}

The colour connections between partons define colour singlet objects, the
clusters. The cluster hadronization model \cite{Webber1984}, which is
implemented in \herwig{}++, generates hadronic final states based on clusters.
Figure~\ref{fig:classesdef} shows that in events with multiple parton scatters
clusters can be discriminated by the origin of their partonic constituents. We
define three classes of clusters. $h$-type clusters consist of partons generated
perturbatively in a single partonic subprocess. The second type of clusters are
the subprocesses-interconnecting ones, which combine partons generated
perturbatively in different subprocesses. These clusters are labelled as
$i$-type. The remaining clusters, which we call $n$-type, contain one parton
which was created non-perturbatively, i.e.~during the extraction of partons from
the hadrons or in soft scatters. Using this classification, we see in
Fig.~\ref{fig:classesmass} that $n$-type clusters contribute most to the
high-mass tail in the invariant mass distribution of the clusters. This
observation is easily interpreted: The non-perturbative extraction of the
partons from the protons, denoted by the grey-shaded area in
Fig.~\ref{fig:classesdef}, may yield colour connections between partons which
are distant in momentum space and thus have large invariant masses. To restore
the physical picture of preconfinement, a colour reconnection model must be
applied which helps to avoid these heavy clusters.

\section{Colour reconnection models}

A colour reconnection model has been included in \herwig{}++ as of version
2.5~\cite{Gieseke:2011na}. This model iterates over all cluster pairs in a
random order. Whenever a swap of colours is preferable, i.e.~when the new
cluster masses are smaller, this is done with a given probability, which is the
only model parameter. This \emph{plain} model has shown to give the desired
results. As the clusters are presented to the model only in a given sequence,
though, it is hard to assess which clusters are affected and to what extent the
sequence is physically relevant.

For these reasons, we implemented another CR model, which adopts the Metropolis
\cite{Metropolis:1953am} and the Simulated-Annealing algorithm
\cite{Kirkpatrick}. The \emph{statistical} colour reconnection model has been
implemented as of \herwig{}++ 2.6 \cite{release26} and is discussed in detail in
Ref.~\cite{Gieseke:CRmodel}.  The new CR model reduces the colour length
$\lambda \equiv \sum_{i=1}^{N_{\mathrm{cl}}} m_{i}^2$\, statistically, where
$N_{\mathrm{cl}}$ is the number of clusters in an event and $m_i$ is the
invariant mass of cluster $i$.

\begin{figure}[t]
  \begin{minipage}[t]{.48\linewidth}
    \centering
    \includegraphics[width=0.8\textwidth]{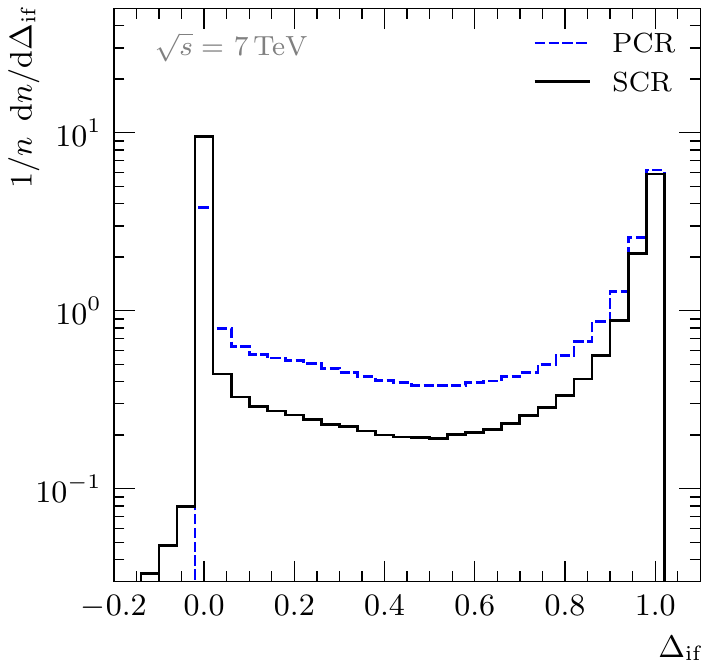}%
    \subcaption{Colour length drop in soft-inclusive $pp$ collisions. PCR
    denotes the plain CR model, whereas SCR stands for the statistical model.}
    \label{fig:colourlengthdrop}
  \end{minipage}%
  \hfill
  \begin{minipage}[t]{.48\linewidth}
    \centering
    \includegraphics[width=0.9\textwidth]{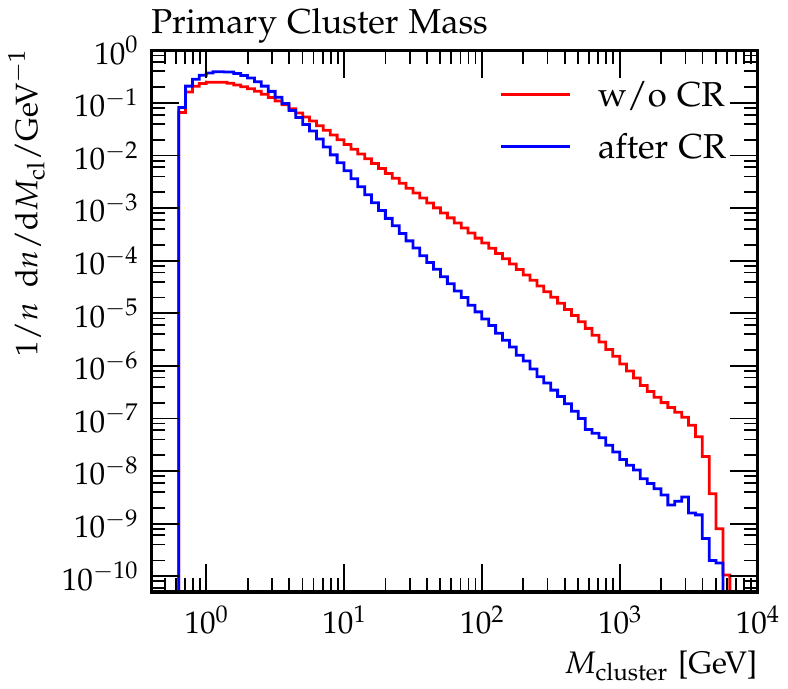}%
    \subcaption{Effect of colour reconnection on the cluster mass spectrum.}
    \label{fig:clustermassshift}
  \end{minipage}
  \caption{Impact of colour reconnection on the colour length and the cluster
  mass spectrum.}
  \label{fig:creffects}
\end{figure}

For both the plain and the statistical CR model we observe an extreme drop in
the colour length, $\deltaif \equiv 1 - \lambdafinal / \lambdainit$, as shown in
Fig.~\ref{fig:colourlengthdrop}. Here, $\lambdainit$ and $\lambdafinal$ denote
the colour length $\lambda$ before and after the colour reconnection procedure,
respectively. The change in the cluster mass spectrum is directly visible in
Fig.~\ref{fig:clustermassshift}. For these plots, a set of typical values for
the model parameters was used, which we obtained from tunes to experimental
data.

\section{Results}

\begin{figure}[t]
  \begin{minipage}[t]{.48\linewidth}
    \centering
    \includegraphics[width=1.0\textwidth]{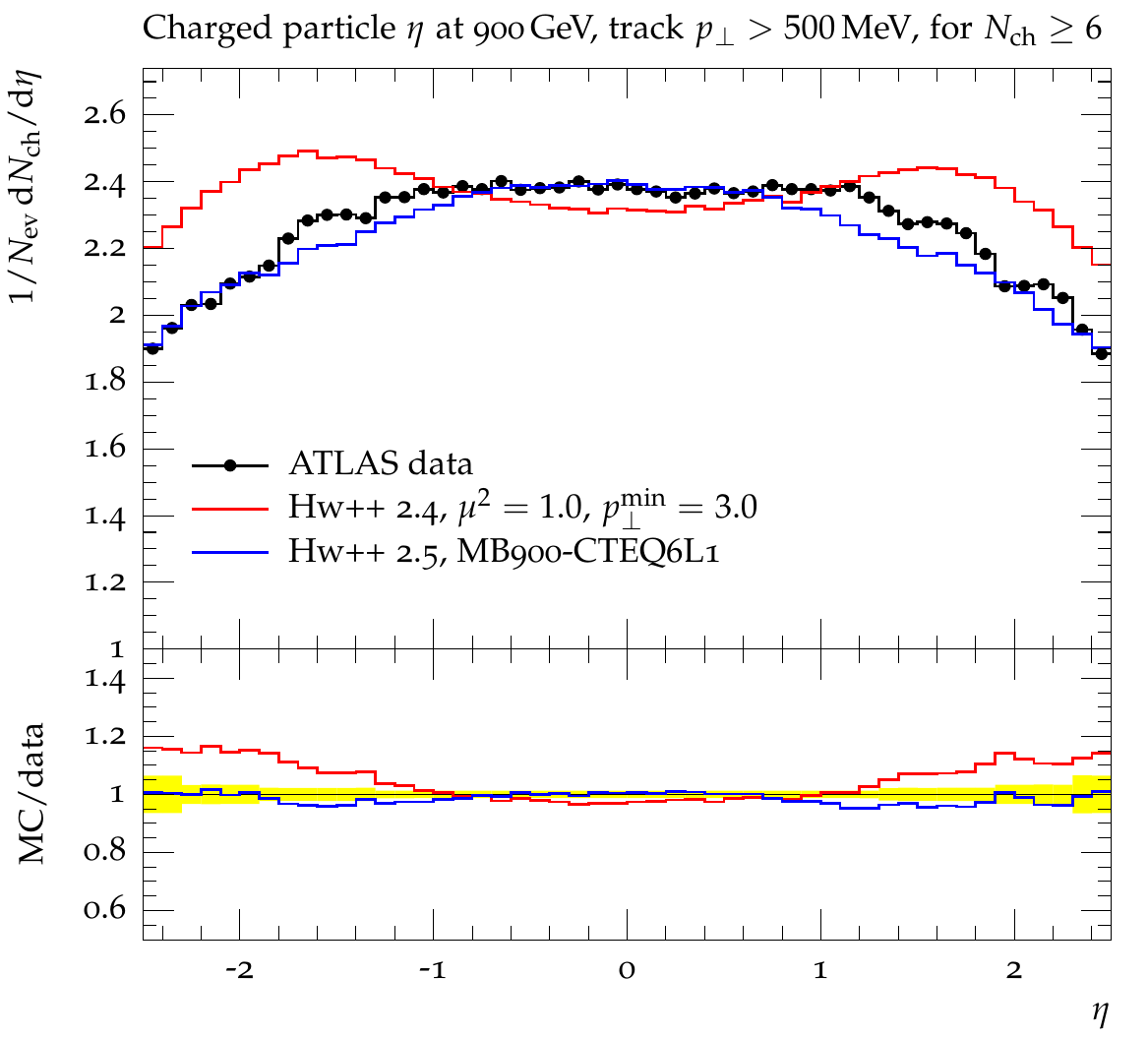}%
    \subcaption{Pseudorapidity distribution of charged particles. The \herwig{}
    2.4 model contains no CR. \textsc{mb900-cteq6l1} is a dedicated tune of the
    model with PCR to 900 GeV MB data.}
    \label{fig:LHC:MB}
  \end{minipage}%
  \hfill
  \begin{minipage}[t]{.48\linewidth}
    \centering
    \includegraphics[width=1.0\textwidth]{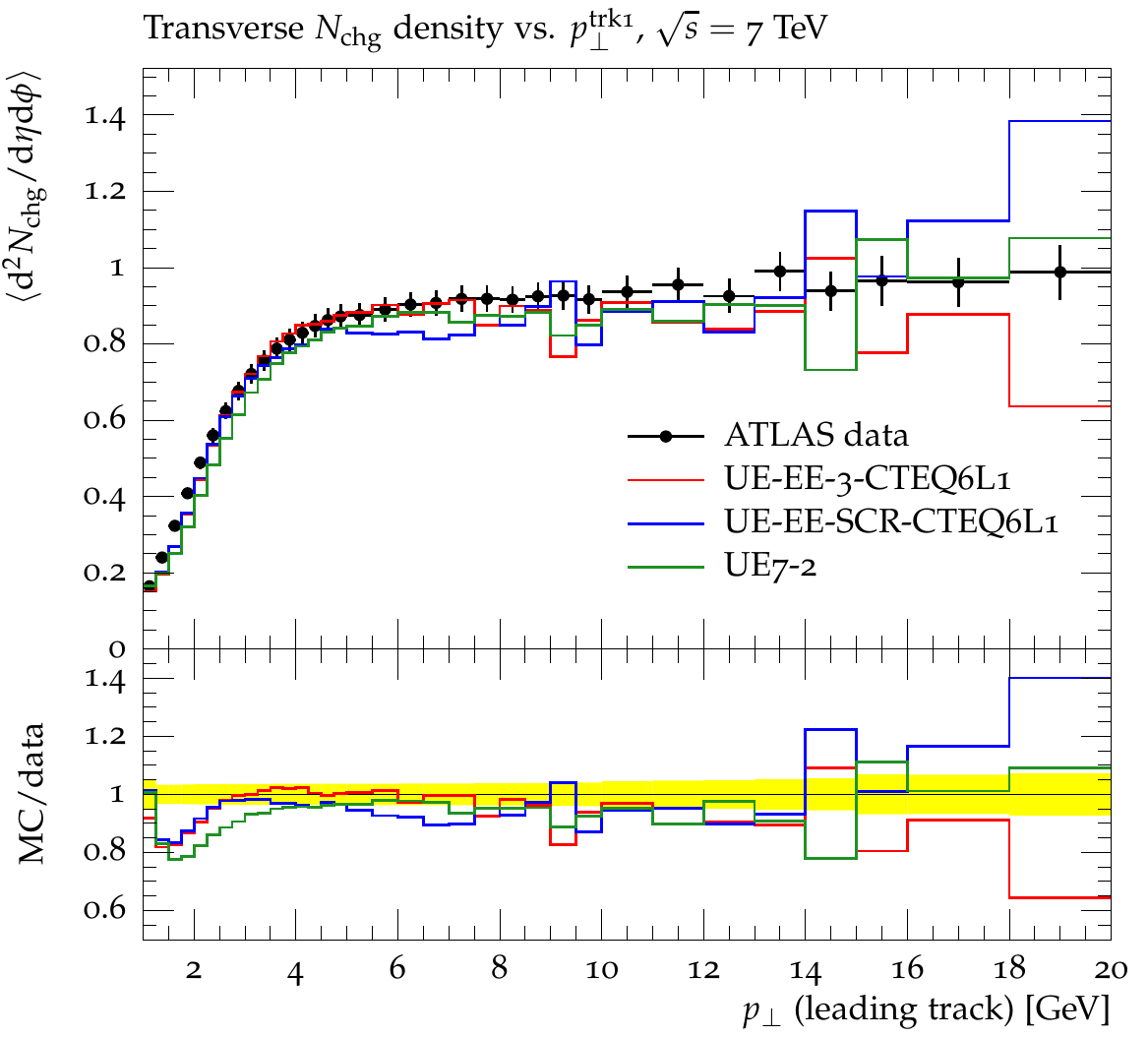}%
    \subcaption{Charged-particle multiplicity density in the transverse area as
    a function of the $p_\perp$ of the leading track.  All histograms show
    \herwig{} UE tunes including CR.}
    \label{fig:LHC:UE}
  \end{minipage}
  \caption{\herwig{} results compared to ATLAS data.}
  \label{fig:LHC}
\end{figure}

We find that CR improves the description of MB data from ATLAS. As an example,
we show in Fig.~\ref{fig:LHC:MB} the pseudorapidity distribution of charged
particles at $\sqrt{s}=\unit{900}{\GeV}$, compared to ATLAS data
from~\cite{Aad2011a}.  This analysis suppresses contributions from diffractive
events by cutting on the transverse momentum of the charged particles, $p_\perp
> \unit{500}{\MeV}$, and on the charged-particle multiplicity, $N_{\rm ch} \geq
6$. As \herwig{}++ contains no model for soft diffraction, a comparison to
samples with looser cuts, $p_\perp > \unit{100}{\MeV}$ and $N_{\rm ch} \geq 1$,
which contain diffractive contributions, yields less agreement. We expand on
this in more detail in~\cite{Gieseke:CRmodel}.

The model also enables a good description of the underlying event. In
Fig.~\ref{fig:LHC:UE} we see, as an example, the charged-particle multiplicity
density at $\sqrt{s}=\unit{7}{\TeV}$, in a region \emph{transverse} to the
leading track in azimuth, $60^\circ < |\Delta \phi| < 120^\circ$, which is most
sensitive to underlying-event activity. The model results are compared to ATLAS
data from \cite{Aad:2010fh}.

\section{Conclusions}

We have summarized the latest developments in the MPI model in \herwig{}++ and
expanded on the motivation and modelling of colour reconnection.  Furthermore,
we have shown that (sufficiently diffraction-suppressed) minimum-bias data from
the LHC and underlying-event observables are well described by the present
model.

\section*{Acknowledgements}

We are grateful to the other members of the \herwig{} collaboration for critical
discussions and support. Moreover, we wish to thank the organizers for this
pleasant workshop.  We acknowledge financial support from the Helmholtz Alliance
``Physics at the Terascale''.


{\raggedright
\begin{footnotesize}
  \bibliographystyle{DISproc}
  \bibliography{rohr_christian.bib}
\end{footnotesize}
}


\end{document}